\documentclass[prd,10pt,twocolumn,showpacs,preprintnumbers,amsmath,amssymb]{revtex4}

\usepackage{graphicx}
\usepackage{dcolumn}
\usepackage{bm}

\begin{document}

\title{Dark energy with polytropic equation of state}

 \author{Utpal Mukhopadhyay}
\affiliation{Satyabharati Vidyapith, Kolkata 700 126, North 24
Parganas, West Bengal, India}

\author{Saibal Ray}
\affiliation{Department of Physics, Barasat Government College,
Kolkata 700 124, North 24 Parganas, West Bengal, India \\and
Inter-University Centre for Astronomy and Astrophysics, PO Box 4,
Pune 411 007, India; e-mail:saibal@iucaa.ernet.in}

\date{\today}

\begin{abstract}
Equation of state parameter plays a significant role for guessing
the real nature of dark energy. In the present paper polytropic
equation of state $p=\omega\rho^n$ is chosen for some of the
kinematical $\Lambda$-models viz., $\Lambda \sim (\dot a/a)^2$,
$\Lambda \sim \ddot a/a$ and $\Lambda \sim \rho$. Although in dust
cases ($\omega=0$) closed form solutions show no dependency on the
polytropic index $n$, but in non-dust situations some new
possibilities are opened up including phantom energy with
supernegative ($\omega<-1$) equation of state parameter.
\end{abstract}


 \maketitle

\section{INTRODUCTION}
Ever since the discovery of an accelerating Universe through SN Ia
observations \cite{rie98,per98}, scientists are searching for the
cause behind this acceleration. It has been suspected that some
kind of yet unknown energy is responsible for injecting the right
amount of energy to the Universe for changing it to an
accelerating one from a decelerating phase. Scientific community
has coined the name {\it dark energy} for this unknown energy.
Various types of models have been proposed for approximating this
dark energy so far \cite{ove98,sah00}.

It is observed that equation of state parameter plays a crucial
role in understanding the actual nature of dark energy
\cite{kir03}. Till now, most of the models of $\Lambda$ (the
so-called cosmological constant of Einstein and a representative
symbol for dark energy) with dynamic character have relied on the
barotropic equation of state with wide range of values of the
equation of state parameter $\omega$, viz., $\omega =0$ (for dust
filled Universe), $\omega =1/3$ (for radiation), $\omega =-1$ (for
vacuum), $-1<\omega<0$ (for quintessence) and $\omega<-1$ (for
phantom energy). But, all these models are plagued by some
shortcomings. For instance, models with $\omega =0$ and $1/3$ show
an excess \cite{vis00} or very low \cite{ove98} age of the
Universe while the stiff-fluid model, in spite of its nice
agreement with the present age of the Universe \cite{ray257}, is
not, in general, accepted as the real nature of the present
Universe. In this circumstances, it is not unreasonable to think
of an equation of state different from the barotropic one. This
has prompted us to investigate the reaction of some of $\Lambda$
models when polytropic equation of state is chosen.

Polytropic equation of state has been used in various
astrophysical situations such as in the case of Lane-Emden models
\cite{tiw86,ray02}. In the present investigation it is used in
cosmological realm. As test models we have selected $\Lambda \sim
(\dot a/a)^2$, $\Lambda \sim \ddot a/a$ and $\Lambda \sim \rho$
models -- the same three kinematical $\Lambda$-models which were
chosen in one of our previous works \cite{ray295} under the
barotropic equation of state. Here Sections 2 and 3 deal with the
field equations and their solutions under three different
$\Lambda$-models while various physical implications of the
present work are discussed in Section 4.\\

\section{EINSTEIN FIELD EQUATIONS}
 The
Einstein field equations are
 \begin{eqnarray}
  R^{ij} - \frac{1}{2}Rg^{ij} = -8\pi G\left[T^{ij} - \frac{\Lambda}{8\pi G}g^{ij}\right]
\end{eqnarray}
where {\it cosmological constant} $\Lambda$ is assumed as a
function of time, viz., $\Lambda = \Lambda(t)$ and the velocity of
light $c$ in vacuum is unity when expressed in relativistic units.

For the spherically symmetric
Friedmann-Lema{\^i}tre-Robertson-Walker (FLRW) metric
\begin{eqnarray}
 ds^2 = -dt^2 + a(t)^2\left[\frac{dr^2}{1 -
kr^2} + r^2 (d\theta^2 + sin^2\theta d\phi^2)\right]
\end{eqnarray}
 where $a$ is the scale factor and $k$, the curvature constant $-1, 0, +1 $
respectively for open, flat and close models of the Universe, the
Einstein field equations (1) take the forms as
\begin{eqnarray} \left(\frac{\dot a}{a}\right)^2 =
\frac{8\pi G}{3}\rho + \frac{\Lambda}{3}, \end{eqnarray}
\begin{eqnarray} \frac{\ddot a}{a} = - \frac{4\pi G}{3} (\rho +
3p)+ \frac{\Lambda}{3}. \end{eqnarray} As various observational
results \cite{ber00} and inflation theory indicate that the
Universe is flat so we have assumed here $k=0$.\\

Let us choose the polytropic equation of state
 \begin{eqnarray}
  p = w {\rho}^{n}
\end{eqnarray}
 where equation of state parameter $w$ can take the constant values
 $0$, $1/3$, $-1$ and $+1$ respectively for the dust, radiation, vacuum
 fluid and stiff fluid and $n$ is the polytropic index.\\

\section{PARTICULAR SOLUTIONS}
\subsection{$\Lambda \sim (\dot a/a)^2$}
 If we use the {\it
ansatz}
\begin{eqnarray}
\Lambda = 3 \alpha\left(\frac{\dot a}{a}\right)^2
 \end{eqnarray}
where $\alpha$ is a constant, then using equation (6) we get from equation (3)

\begin{eqnarray}
3 \left(\frac{\dot a}{a}\right)^2- 3 \alpha\left(\frac{\dot a}{a}\right)^2= 8 \pi G \rho
 \end{eqnarray}
which takes the form for density as follows:
\begin{eqnarray}
\rho = \frac{3(1- \alpha)}{ 8 \pi G}{\left(\frac{\dot a}{a}\right)^2}.
 \end{eqnarray}
Using equation (5) and equation (6) we get from equation (4)
\begin{eqnarray}
3 \left(\frac{\ddot a}{a}\right) = -12 \pi G\omega \rho^n- 4\pi G \rho+3 \alpha\left(\frac{\dot a}{a}\right)^2
\end{eqnarray}

which, on simplification, yields
\begin{eqnarray}
a^{2n-1}\frac{d^{2}a}{dt^2} + A \left(\frac{da}{dt}\right)^{2n} + B a^{2n-2}\left(\frac{da}{dt}\right)^2=0
\end{eqnarray}
where $A=[3^{n} \omega(1-\alpha)^n]/[2^{n}(4 \pi G)^{n-1}]$ and $B=(1-3\alpha)/2$.

Let us substitute, $z=da/dt$ for which the equation (10) reduces to
\begin{eqnarray}
z^{1-2n}\frac{dz}{da}+Aa^{1-2n}+z^{2-2n}\frac{B}{a}=0.
\end{eqnarray}
If we put, $z^{2-2n}=y$, then differentiating it with respect to the scale factor $a$, we get
\begin{eqnarray}
z^{1-2n}\frac{dz}{da}=\frac{1}{(2-2n)}\frac{dy}{da}.
\end{eqnarray}
Hence equation (11), by use of equation (12) becomes,
\begin{eqnarray}
\frac{dy}{da}+\frac{(2-2n)B}{a}y=-(2-2n)Aa^{1-2n}.
\end{eqnarray}
Now let us solve equation (13) in some particular cases for getting insight into the physical situations.\\

\subsubsection{Dust case ($\omega =0$)}
 For pressureless dust,
$\omega=0=A$ and equation (13) becomes,
\begin{eqnarray}
\frac{dy}{da}+\frac{(2-2n)B}{a}y=0.
\end{eqnarray}
Solving equation (14) we get,
\begin{eqnarray}
y=C_{1}a^{(2n-2)B}
\end{eqnarray}
$C_1$ being an integration constant.

This immediately yields
\begin{eqnarray}
\frac{da}{dt}=C_2 a^{-B}
\end{eqnarray}
where $C_2={C_{1}}^{1/(2-2n)}$. \\
Integrating equation (16) and
using the initial condition that $a=0$ for $t=0$ we get,
\begin{eqnarray}
a(t)=C_3\left[\frac{3(1-\alpha)}{2}\right]^{2/3(1-\alpha)}t^{2/3(1-\alpha)}
\end{eqnarray}
where $C_3= {C_2}^{1/(B+1)}$ being an integration constant, which
provides the Hubble parameter as
\begin{eqnarray}
H(t)=\frac{\dot a}{a}=\frac{2}{3(1-\alpha)}\frac{1}{t}.
\end{eqnarray}
Therefore from (8), we get
\begin{eqnarray}
\rho(t)=\frac{1}{6\pi G(1-\alpha)}\frac{1}{t^2}.
\end{eqnarray}
Also, from equation (6), we obtain
\begin{eqnarray}
\Lambda(t)=\frac{4\alpha}{3(1-\alpha)^{2}}\frac{1}{t^2}.
\end{eqnarray}
Thus, it is interesting to note that in dust case, expressions for
$\rho(t)$ and $\Lambda(t)$ as well as $t$ dependent part of $a(t)$
are independent of $n$. Moreover, $a(t)$, $\rho(t)$ and
$\Lambda(t)$ follow the same power law with $t$ as those in the
dust case with barotropic equation of state (i.e. $n=1$)
\cite{ray295}. But, surprisingly, equation (17) becomes undefined
for  $n=1$ since $C_3$ contains the factor $(2-2n)$ in the
denominator.  \\

\subsubsection{Non-dust case ($\omega \neq 0$)}
 When $\omega\neq 0$, then $ A\neq
0$ and hence equation (11) is a linear equation. Multiplying
equation (11) by integrating factor $ a^{(2-2n)B} $ and solving
the resulting equation we get,
\begin{eqnarray}
y=-\frac {Aa^{(2-2n)B}}{B+1}
\end{eqnarray}
which yields after substituting the value of $y$
\begin{eqnarray}
\frac {da}{dt}=-\frac{A}{(B+1)}a.
\end{eqnarray}
By solving equation (22) we get our solution for the scale factor in the form
\begin{eqnarray}
a(t)=C_4 exp [(-\omega)^{1/2(1-n)}\frac {8\pi G}{3(1-\alpha)}]^{1/2}t
\end{eqnarray}
where $C_4$ is integration constant.\\
 It is clear from equation
(23) that no real value of $a(t)$ is possible if $\omega$ is
positive. So, $\omega$ must be negative. Putting specifically,
$\omega=-1$ in equation (23) we obtain
\begin{eqnarray}
a(t)=C_4 exp [8 \pi G/3(1-\alpha)]^{1/2}t.
\end{eqnarray}
Now, $\omega=-1$ means a vacuum fluid. So, for vacuum fluid, we get an exponential solution
independent of $n$. But, if $\omega<-1$, then scale factor depends on $n$.
Now, $\omega<-1$ corresponds to a supernegative equation of state and hence the idea of
phantom energy comes into the picture. Also, during inflation, the  Universe
underwent an exponential expansion. So, for non-dust case our solution reflects an
inflationary scenario which may be due to either vacuum energy or quintessence
or phantom energy.
Also, from equation (24) we get,
\begin{eqnarray}
H=\left[ \frac{8 \pi G}{3(1-\alpha)}\right]^{1/2}.
\end{eqnarray}
From equation (8) and (6) we get respectively
\begin{eqnarray}
\rho(t)=1,
\end{eqnarray}
\begin{eqnarray}
\Lambda(t)=\frac{8 \pi G \alpha}{1-\alpha}.
\end{eqnarray}
Therefore, the cosmic matter and vacuum energy densities are given by
\begin{eqnarray}
\Omega_m=\frac{8 \pi G \rho}{3H^2}=1-\alpha,
\end{eqnarray}
\begin{eqnarray}
\Omega_{\Lambda}=\frac{\Lambda}{3H^2}=\alpha.
\end{eqnarray}
So that, equations (29) and (30), immediately provide
\begin{eqnarray}
\Omega_m+\Omega_{\Lambda}=1.
\end{eqnarray}
Thus, we find that although polytropic equation of state yields
interesting result by invoking the idea of phantom energy for
non-dust case, yet it presents us some unacceptable situations
like constant $H$ (equation (25)) and constant $\rho$ (equation
(26)). A possible explanation of this will be discussed
afterwards. However, the cosmic matter and vacuum energy density
parameters satisfy the well-known relation $\Omega_m+\Omega
_{\Lambda}=1$ for flat Universe ($\Omega_k=0$). \\

\subsection{$\Lambda \sim \rho$}
 For this model we use
the {\it ansatz}
\begin{eqnarray}
\Lambda=8 \pi G\gamma\rho
\end{eqnarray}
where $\gamma$ is a constant.\\
Then, from equation (3) we get
\begin{eqnarray}
\rho=\frac{3}{8 \pi G(\gamma+1)}\left( \frac{\dot a}{a}\right)^2.
\end{eqnarray}
Using (5) we get from (4)
\begin{eqnarray}
3\left(\frac{\ddot a}{a}\right)=-12\pi G\omega\rho^n-4\pi G\rho+8\pi G\gamma\rho
\end{eqnarray}
which, on simplification, yields the differential equation
\begin{eqnarray}
a^{(2n-1)}\frac{d^{2}a}{dt^{2}}+D
\left(\frac{da}{dt}\right)^{2n}-Ea^{(2n-2)}\left(\frac{da}{dt}\right)^2=0
\end{eqnarray}
where $D=\omega 3^{n}/[2^{n}(\gamma+1)^{n}(4\pi G)^{n-1}]$ and
$E=(2\gamma-1)/2(\gamma+1)$.\\
Substituting $da/dt=u$ and then
$v=u^{2-2n}$ equation (34) becomes
\begin{eqnarray}
\frac {1}{(2-2n)}\frac {dv}{da}-\frac {E}{a}v=-Da^{1-2n}.
\end{eqnarray}

\subsubsection{Dust case ($\omega=0$)}
 For pressureless dust case $D$=0, so that equation (35) reduces to
\begin{eqnarray}
\frac{dv}{da}+(2n-2)\frac{E}{a}v=0.
\end{eqnarray}
Solving equation (36) in the same manner as in the previous model,
we get our solution set as
\begin{eqnarray}
a(t)=C_5 \left[\frac
{3}{2(\gamma+1)}\right]^{2(\gamma+1)/3}t^{2(\gamma+1)/3}
\end{eqnarray}
where $C_5$ is a constant.
\begin{eqnarray}
 H(t)=\frac{2(\gamma+1)}{3}\frac{1}{t},
\end{eqnarray}
\begin{eqnarray}
\rho(t)=\frac{(\gamma+1)}{6 \pi G}\frac{1}{t^2},
\end{eqnarray}
\begin{eqnarray}
\Lambda(t)=\frac{4\gamma(\gamma+1)}{3}\frac{1}{t^2}.
\end{eqnarray}
Also, we can calculate for the cosmic and vacuum energy densities
which, respectively, are $\Omega_m=1/(\gamma+1)$ and
$\Omega_{\Lambda}=\gamma/(\gamma+1)$. Therefore,
$\Omega_m+\Omega_{\Lambda}=1$ and hence
$\gamma=\Omega_{\Lambda}/\Omega_m$. Thus, we find that for this
model also $a(t)$, $\rho(t)$ and $\Lambda(t)$ follow the same
relationship with time as in the barotropic case \cite{ray295} and
$\gamma$ is related to matter and vacuum energy densities by the
same relation as in the barotropic case \cite{ray295}.

\subsubsection{Non-dust case ($\omega \neq 0$)}
 When $\omega \neq 0$, then $D
\neq 0$. In this case, following the same procedure as in the
$\Lambda \sim (\dot a/a)^{2}$ model, it is easy to show that no
real solution is possible for  $\omega>0$.\\ For $\omega=-1$, we
have
\begin{eqnarray}
a(t)=C_6 exp \left[\frac{3}{8\pi G(\gamma+1)}\right]^{1/2}t
\end{eqnarray}
where $C_6$ is a constant.\\
\begin{eqnarray}
H=\left[\frac{3}{8\pi G(\gamma+1)}\right]^{1/2},
\end{eqnarray}
\begin{eqnarray}
\rho(t)=\left[\frac{3}{8\pi G(\gamma+1)}\right]^{2},
\end{eqnarray}
\begin{eqnarray}
\Lambda(t)=\frac{9\gamma}{8\pi G(\gamma+1)^{2}}.
\end{eqnarray}
Here we find that in non-dust case, solutions are possible for
vacuum fluid ($\omega=-1$), quintessence ($-1<\omega<0$) and
phantom energy ($\omega<-1$). For vacuum fluid, solution is
independent of $n$ whereas for quintessence and phantom energy
solution depends on $n$. In this case also, $\gamma$ is found to
be the ratio of vacuum energy density and matter energy density.
For physical reality, $\gamma>-1$ to be imposed on the solutions.

\subsection{$\Lambda \sim \ddot a/a$}
If we use the supposition $\Lambda=\beta(\ddot a/a)=\beta(\dot
H+H^{2})$, then from equation (3) we get,
 $3H^{2}=8\pi G\rho+\beta(\dot H+H^{2})$. Therefore
\begin{eqnarray}
\rho=\frac{1}{8\pi G}[(3-\beta)H^{2}-\beta \dot H].
\end{eqnarray}
Then, using (5), from (4) we have, $3(\dot H+H^{2})=-12\pi G\omega\rho^{n}-
4\pi G\rho+\beta(\dot H+H^{2})$ which on simplification becomes
\begin{widetext}
\begin{eqnarray}
\frac{3}{2}\left[(3-\beta)H^{2}+ (2-\beta)\dot H \right]
=-\frac{12\pi G\omega}{(8\pi G)^n} [(3-\beta)H^{2}-\beta \dot
H]^{n}.
\end{eqnarray}
\end{widetext}
For simplifying equation (46), let us investigate in
some particular cases. If we specifically choose $n=0$ (i.e.
$p=\omega$), then equation (46) reduces to
\begin{eqnarray}
(2-\beta)\dot H+(3-\beta)H^2=-8\pi G\omega.
\end{eqnarray}

\subsubsection{Dust case ($\omega=0$)}
For pressureless dust ($\omega=0$) equation (47) becomes
\begin{eqnarray}
(2-\beta)\frac{dH}{dt}+(3-\beta)H^2=0.
\end{eqnarray}
By solving equation (48) and using the initial condition that $a=0$ when $t=0$ we get
\begin{eqnarray}
\frac{da}{dt}= a \left(\frac{2-\beta}{3-\beta}\right)\frac{1}{t}.
\end{eqnarray}
Then from equation (49) we finally get our solution set as
\begin{eqnarray}
a(t)=C_7 t^{(\beta-2)/(\beta-3)}
\end{eqnarray}
where $C_7$ is a constant.\\
\begin{eqnarray}
\rho(t)=\frac{1}{4\pi
G}\left(\frac{\beta-2}{\beta-3}\right)\frac{1}{t^2},
\end{eqnarray}
\begin{eqnarray}
\Lambda(t)=\frac{\beta(\beta-2)}{(\beta-3)^2}\frac{1}{t^2},
\end{eqnarray}
\begin{eqnarray}
H(t)=\left(\frac{\beta-2}{\beta-3}\right)\frac{1}{t}.
\end{eqnarray}
Thus, for $n=0$, the solution set is same for $a(t)$, $\rho(t)$
and $\Lambda(t)$ as obtained by Ray and Mukhopadhyay \cite{ray295}
in the dust case with barotropic equation of state.

\subsubsection{Non-dust case ($\omega \neq 0$)}
In non-dust case, equation (47) can be written as
\begin{eqnarray}
(2-\beta)\frac{dH}{dt}+(3-\beta)H^2=-8\pi G\omega.
\end{eqnarray}
By solving equation (47) and remembering that $H=0$ when $t=0$ we get
\begin{eqnarray}
\frac{da}{dt}=\nu a \quad tan(\mu t)
\end{eqnarray}
where $\mu=[8\pi  G\omega(3-\beta)/(\beta-2)]^{1/2}$ and $\nu=[8\pi  G\omega/(3-\beta)]^{1/2}$.\\
After solving equation (55) we have
\begin{eqnarray}
a(t)=C_8\left[sec \frac{\{8\pi
G\omega(3-\beta)\}^{1/2}}{(\beta-2)}t\right] ^{\{8\pi
G\omega/(3-\beta)\}^{1/2}},
\end{eqnarray}
\begin{widetext}
\begin{eqnarray}
\rho(t)=\frac{(8\pi G\omega)^{3/2}(3-\beta)^{1/2}}{8\pi G
(\beta-2)^2} \left[\{8\pi G\omega(3-\beta)\}^{1/2} tan^2
\frac{\{8\pi  G\omega(3-\beta)\}^{1/2}}{(\beta-2)}t- \beta sec^2
\frac{\{8\pi G\omega(3-\beta)\}^{1/2}}{(\beta-2)}t\right],
\end{eqnarray}
\end{widetext}
\begin{widetext}
\begin{eqnarray}
\Lambda(t)=\frac{(8\pi
G\omega)^{3/2}}{(\beta-2)^2}\beta\left[(3-\beta)^{1/2} sec^2
\frac{\{8\pi  G\omega(3-\beta)\}^{1/2}}{(\beta-2)}t +(8\pi
G\omega)^{1/2} tan^2  \frac{\{8\pi
G\omega(3-\beta)\}^{1/2}}{(\beta-2)}t\right],
\end{eqnarray}
\end{widetext}
\begin{eqnarray}
H(t)=\frac{8\pi  G\omega}{(\beta-2)}tan  \frac{\{8\pi
G\omega(3-\beta)\}^{1/2}}{(\beta-2)}t.
\end{eqnarray}
From the above solution set, it is clear that if $\omega>0$, then for real $a(t)$, $\rho(t)$, $\Lambda(t)$ and
$H(t)$, $\beta$ must be less than $3$. For negative $\omega$ (i.e., for vacuum fluid, quintessence and phantom energy)
real values of $a(t)$ and $H(t)$ can be obtained, but $\Lambda(t)$ and $\rho(t)$ become imaginary. If $(\beta-3)=0$,
then $\rho(t)$, $\Lambda(t)$ and $H(t)$ are zero but $a(t)$ is undefined. Since $tan t$ and $sec t$ are
 both increasing functions of $t$, then $a(t)$, $\Lambda(t)$ and $H(t)$ increase with time. Increasing $a(t)$ supports
the idea of an accelerating Universe, but increasing $\Lambda(t)$ and $H(t)$ are contrary to the present status of those two parameters.\\
For $t=0$, we get respectively from equations (56) - (59) the following physical parameters:
\begin{eqnarray}
a(t)=C_8,
\end{eqnarray}
\begin{eqnarray}
\rho(t)=-\beta \frac{1}{8\pi G}\frac{(8\pi
G\omega)^{3/2}(3-\beta)^{1/2}}{(\beta-2)^2},
\end{eqnarray}
\begin{eqnarray}
\Lambda(t)=\frac{(8\pi  G\omega)^{3/2}}{(\beta-2)^2}\beta
(3-\beta)^{1/2},
\end{eqnarray}
\begin{eqnarray}
H(t)=0.
\end{eqnarray}
Since $a(t)$ assumes a constant value for $t=0$, then our model hints at the existence of some kind of quantum fluctuation at the time of Big Bang.
Now, equation (61) tells us that for physically valid $\rho$, $\omega>0$ and $\beta<0$. For a negative $\beta$ we get an attractive $\Lambda$.
So, as a whole this model also presents some interesting as well as awkward cosmological picture.\\

\section{DISCUSSION}
 Present investigation reveals that for
pressureless dust, all the three models reflect the same result as
obtained by Ray and Mukhopadhyay \cite{ray295} in dust case with
barotropic equation of state. But, in non-dust cases (i.e.,
$\omega \neq 0$), $\Lambda \sim (\dot a/a)^2$ and $\Lambda \sim
\rho$ models show exponential expansion of the Universe for
negative $\omega$ whereas no real situation is possible for
$\omega>0$. Thus, we can say that for dust-filled Universe, there
is no distinction between barotropic and polytropic equations of
state. This is not at all unexpected because for $\omega=0$, we
have $p=0$, whatever may be the value of $n$ in the equation of
state. Also, non-dust cases for all the three models present us
some unpleasant results in terms of constant $H$ and $\rho$. These
situations can be explained if we assume that the non-dust cases
reflect the picture of early Universe when inflation occurred due
to the presence of quintessence ($-1<\omega<0$) or vacuum fluid
($\omega=-1$) or phantom energy ($\omega<-1$). So, using
polytropic equation of state it has been possible to show that
non-dust cases admit the presence of a driving force behind
inflation in the form of either quintessence or vacuum fluid or
phantom energy and in the dust cases there is no distinction
between different equation of states. Moreover, present models do
not depend on any particular value of $n$ in the polytropic
equation of state. This proves the generality of the present
investigation.\\

\begin{acknowledgments} One of the authors (SR) would like to
express his gratitude to the authority of IUCAA, Pune for
providing him the Associateship Programme under which a part of
this work was carried out.
\end{acknowledgments}

\end{document}